\documentclass[prd,aps,preprint,draft,showpacs,floatfix]{revtex4-1}
\usepackage{epsf}
\usepackage[dvips]{color}
\usepackage{dcolumn}
\newcommand{\rSXPT}{rS$\chi$PT}
\newcommand{\Tr}{{\rm Tr}}
\newcommand{\ket}[1]{\left|#1\right\rangle}
\newcommand{\ie}{{\it i.e.}}
\newcommand{\Dslash}{{D}\!\!\!\!/\,}

\begin{document}
\title{Topological susceptibility with the asqtad action}
\author{
A.~Bazavov and D.~Toussaint
}
\affiliation{
Department of Physics, University of Arizona, Tucson, AZ 85721, USA
}
\author{C.~Bernard and J.~Laiho
\footnote{Present address, University of Glasgow, Glasgow G12 8QQ, UK}
}
\affiliation{
Department of Physics, Washington University, St.~Louis, MO 63130, USA
}
\author{
B.~Billeter, C.~DeTar, L.~Levkova, and M.B.~Oktay
}
\affiliation{
Physics Department, University of Utah, Salt Lake City, UT 84112, USA
}
\author{
Steven Gottlieb
}
\affiliation{
Department of Physics, Indiana University, Bloomington, IN 47405, USA\\
NCSA, University of Illinois, Urbana IL 61801, USA
}
\author{
U.M.~Heller
}
\affiliation{
American Physical Society, One Research Road, Ridge, NY 11961, USA
}
\author{
J.E.~Hetrick
}
\affiliation{
Physics Department, University of the Pacific, Stockton, CA 95211, USA
}
\author{
J.~Osborn
}
\affiliation{
Argonne National Laboratory, Argonne, IL, USA
}
\author{
R.L.~Sugar
}
\affiliation{
Department of Physics, University of California, Santa Barbara, CA 93106, USA
}
\author{
R.S.~Van de Water
}
\affiliation{
Department of Physics, Brookhaven National Laboratory, Upton, NY 11973, USA
}
\author{(MILC Collaboration)}\noaffiliation
%
\date{\today}

\begin{abstract}
Chiral perturbation theory predicts that in quantum chromodynamics
(QCD), light dynamical quarks suppress the gauge-field topological
susceptibility of the vacuum.  The degree of suppression depends on
quark multiplicity and masses.  It provides a strong consistency test
for fermion formulations in lattice QCD\@.  Such tests are especially
important for staggered fermion formulations that lack a full chiral
symmetry and use the ``fourth-root'' procedure to achieve the desired
number of sea quarks.  Over the past few years we have measured the
topological susceptibility on a large database of 18 gauge field
ensembles, generated in the presence of $2+1$ flavors of dynamical
asqtad quarks with up and down quark masses ranging from 0.05 to 1 in
units of the strange quark mass and lattice spacings ranging from
0.045 fm to 0.12 fm.  Our study also includes three quenched ensembles
with lattice spacings ranging from 0.06 to 0.12 fm. We construct the
topological susceptibility from the integrated point-to-point
correlator of the discretized topological charge density $F \widetilde
F$.  To reduce its variance, we model the asymptotic tail of the
correlator.  The continuum extrapolation of our results for the
topological susceptibility agrees nicely at small quark mass with the
predictions of lowest-order SU(3) chiral perturbation theory, thus
lending support to the validity of the fourth-root procedure.
\end{abstract}

\maketitle

\section{Introduction}
\label{sec:intro}

The rich topological structure of the QCD vacuum is known to be
responsible for many interesting nonperturbative effects, such as the
chiral anomaly and chiral symmetry breaking, instantons, and the large
mass of the $\eta^\prime$ meson. Among the wide variety of ways of
looking at these phenomena, one may consider the effect that
topological charge has on the kernel of the Dirac operator.  It has
broad implications. For example, it is intimately
connected with the value of the chiral condensate
\cite{Leutwyler:1992yt}.

The topological susceptibility $\chi_t$ characterizes the tunneling
rate between topologically distinct vacua by instantons and shows up
in low energy phenomenology through the Witten-Veneziano formula
\cite{Witten:1979vv,Veneziano:1979ec} and in chiral perturbation
theory. A gauge configuration with topological charge $\nu$ requires
at least $\nu$ fermionic zero-modes of the Dirac operator. The effect
of quark mass on the topological susceptibility can be seen by
separating the fermion determinant for a particular gauge field
configuration into zero and non-zero modes. For $N_f$ flavors we have
\cite{Vafa:1984xh,Leutwyler:1992yt}
\begin{equation}
\prod_{f=1}^{N_f} \det(\Dslash + m_f)
   = \prod_{f=1}^{N_f}\left[m_f^{|\nu|} \prod_{\lambda>0} 
  (\lambda^2 + m_f^2)\right],
\label{eq:det}
\end{equation}
where $\lambda$ is the imaginary part of the eigenvalue of $\Dslash$.
Thus gauge configurations of nontrivial topology tend to be suppressed
as any one of the quark masses approaches zero.  However, this effect
is compensated at increasing volume by a growing tendency of gauge
field fluctuations to produce nontrivial topology.  Chiral
perturbation theory tells us \cite{Leutwyler:1992yt} that the outcome
of the competition is controlled by the parameter $x = V \Sigma
m^\prime$, where $\Sigma$ is the chiral condensate, $V$ is the
Euclidean space-time volume, and $m^\prime$ is the reduced mass
\begin{equation}
  1/m^\prime = 1/m_1 + 1/m_2 +\ldots{} \, .
\end{equation}
When at least one quark mass gets small at fixed volume (the
``epsilon'' regime, $x \ll 1$), gauge configurations with nontrivial
topological charge are strongly suppressed, as implied by
Eq.~(\ref{eq:det}).  In the physical regime, in which $x \gg 1$,
which is the case for our study, topologically nontrivial
configurations are not suppressed.  Instead, leading order chiral
perturbation theory requires that the mean squared topological charge
be equal to the parameter $x$:
\begin{equation}
  \langle \nu^2 \rangle = V\Sigma m^\prime,
\end{equation}
where the angle brackets represent an average over gauge fields.  Thus
the topological susceptibility,
\begin{equation}
\chi_t = \langle \nu^2 \rangle/V = \Sigma m^\prime,
\label{eq:def_topo_susc}
\end{equation}
remains finite in the large-volume limit.  Even so, it is still
suppressed as $m^\prime \to 0$.

While lattice simulations of QCD have enjoyed considerable success in
recent years, with errors on hadronic spectroscopy computations at the
1--2\% level, simulations have struggled to reproduce this dependence
of $\chi_t$ on both $m_f$ and $N_f$, until recently.  This progress
has come with improvements in lattice fermion technology, which has
given much more control over chiral symmetry and lattice artifacts.

In this article we present results for the dependence of $\chi_t$ on
the quark mass (through the taste-singlet pion mass) using improved
staggered fermions (asqtad formulation). Descriptions of the asqtad
formulation have been given elsewhere \cite{Bazavov:2009bb}.  To
eliminate contributions from unwanted fermion doublers, the staggered
formulation takes the fourth root of the fermion determinant
$\sqrt[4]{\det [\Dslash + m_f]}$ for each quark (``fourth-root
procedure''), which may raise questions about flavor counting.
For a discussion of the issues, please see
\cite{Bazavov:2009bb} and references therein. The
primary purpose of our study, then, is to test the ability of the
fourth root procedure to produce the correct number of sea quarks.
Since the topological susceptibility is measured directly on the gauge
field configuration without the involvement of valence quarks, it is
directly sensitive to sea quark effects.  We will show that the
continuum extrapolation of our results agrees well with lowest-order
SU(3) chiral perturbation theory.

This article summarizes results of calculations carried out over the
past few years on ensembles with $(2+1)$ flavors of asqtad quarks as
they were being generated (see the Appendix). We continue
to use the methodology of our previous work at larger lattice spacing
and quark mass \cite{Bernard:2003gq,Aubin:2004qz,Bernard:2007ez} with
some refinements which appear here. The key features of our approach
are these:

\begin{enumerate}
\item 

obtaining the square of the topological charge from the integral of
the two-point correlator of the topological charge density.

\item

reducing the variance of the integral by modeling the asymptotic
form of the correlator in terms of known hadronic contributions,
and

\item
analyzing the quark-mass and lattice-spacing dependence of the
resulting susceptibility in terms of predictions from rooted
staggered chiral perturbation theory.

\end{enumerate}

In the following section, we discuss the details of our method for
calculating the topological susceptibility on the lattice. We present
our results and analysis in Sec.~\ref{sec:results}.  Finally, we
comment on our results in the Conclusions, Sec.~\ref{sec:conclusions}.
The Appendix lists the parameters of the gauge field ensembles used in
this study.

\section{Methodology}

\label{sec:methods}

\subsection{Definition of the topological susceptibility}

We introduced the topological susceptibility in
Eq.~(\ref{eq:def_topo_susc}) as the mean squared charge per unit
volume: $\chi_t = \langle \nu^2 \rangle/V$. The net topological charge
$\nu$ is the integral over Euclidean space-time of the topological
charge density,
\begin{equation}
   \rho(x) = \frac{1}{32\pi^2} F^a_{\mu\nu} \widetilde F^a_{\mu\nu}.
\end{equation}
The susceptibility is then given by the integral of the correlator of
the charge density, provided the integral is well defined.
\begin{equation}
  \chi_t = \int d^4x \, C(r) \ \ \ \mbox{with} \ \ \  C(r) = \langle \rho(x) \rho(0) \rangle \ ,
\label{eq:corrint}
\end{equation}
where $r = |x|$.  Because the exponential decay of the correlator at
large $r$ is set by nonzero hadron masses, we see that the
susceptibility is properly regarded as a local observable, 
{\it i.e.}, it can be defined in terms of a correlator that has finite
physical range.  We use this definition of the susceptibility,
coupled with a smeared lattice discretization of $F \widetilde F$.

In the continuum limit the integral definition above is problematic.
The unregulated correlator $C(r)$ is nonintegrable: it has a positive,
divergent contact term (at the origin) and, close to the origin, a
compensating negative ultraviolet singularity of order (up to possible
logarithms) $r^{-8}$ \cite{DiVecchia:1981qi, Shuryak:1994rr,
Giusti:2001xh, Seiler:2001je}.  Cancellation is required in order to
produce the expected finite integral Eq.~(\ref{eq:def_topo_susc}).  To
circumvent this mathematical difficulty L\"uscher formulated a
definition of the topological susceptibility in terms of a product of
pseudoscalar and scalar densities of Ginsparg-Wilson quarks
\cite{Luscher:2004fu}.  Since the definition requires computing
all-to-all quark-line disconnected correlators, it is more difficult
to implement, and, to our knowledge, it has not yet been put into
practice.

For present purposes we resort to the naive definition in
Eq.~(\ref{eq:corrint}) and trust that the lattice cutoff and a
smoothed definition of $\rho(x)$ regulate the compensating
singularities enough over a range of reasonably small lattice spacings
that we can test the expected suppression of the topological
susceptibility.  In our present scheme we fix the smoothing scale in
lattice units as we take the lattice spacing to zero.  Our numerical
simulation provides a practical test of the limitations of such a
scheme.  If it fails, as the lattice spacing is decreased, we would
expect to encounter a growing variance from contributions to the
integral near the origin.  This would not invalidate the
method:  The central value in the continuum limit is finite even if
the variance is unbounded.  It could, however, require an
impractically large computational effort to achieve a desired accuracy
as we make the lattice spacing smaller.  We return to this question
after presenting our results.

There are a variety of lattice methods for obtaining the topological
charge.  The traditional ``algebraic'' method uses a lattice
discretization of the density $F \widetilde F$, constructed at each
lattice site from appropriate closed loops of gauge links.  To
suppress ultraviolet noise at the cutoff scale, smoothing is required
\cite{Teper:1985ek}.  The Boulder discretization
\cite{DeGrand:1997gu,Hasenfratz:1998qk}, which we use in the present
study, is a refinement of the traditional definition. It is fattened
(smoothed) by first performing some number (we use three) of HYP
smoothing sweeps \cite{Hasenfratz:2001hp} on the gauge field and then
constructing the operator from the smoothed links.

A more elegant method defines the topological charge density in terms
of a chiral (e.g.\ overlap) Dirac operator $D$, as $\rho(x) \propto
\Tr [\gamma_5 D]_{x,x}$ \cite{Hasenfratz:1998ri,Luscher:1998pqa} 
(the trace is over
color and spin), but using it directly in Eq.~(\ref{eq:corrint}) is
computationally expensive \cite{DelDebbio:2004ns,Horvath:2005cv}.  For
the overlap operator a more tractable method uses the Atiyah-Singer
index theorem to relate the topological charge $\nu$ to the net number
of zero crossings of the low-lying eigenvalues of a Hermitian Dirac
kernel from which the chiral operator is built
\cite{Narayanan:1994gw}.  This method was implemented in
\cite{Chiu:2008jq}. For the overlap operator, smoothing is inherent in
the choice of the Dirac kernel from which the overlap action is built.

Another promising method works with gauge configurations
of fixed topology \cite{Fukaya:2004kp,Aoki:2007ka}.  In this case, at
large distance the correlator of the topological charge density
approaches a constant $\chi_t/V$ plus other known constants that
depend on the fixed topological charge.  One can also use a hadronic
flavor-singlet interpolating operator with $J^{PC} = 0^{-+}$ as a
proxy for $F \widetilde F$.  This method has been tested at one
lattice spacing in the two-flavor case on configurations generated
with the overlap action \cite{Aoki:2007pw}.

The L\"uscher definition \cite{Luscher:2004fu}, based on a
  chiral Dirac operator, replaces the integral of $F \widetilde F$
  with the integral of a quark pseudoscalar density.  The quark field
  from which that density is constructed can have arbitrary mass,
  which sets the localization scale of the operator.  The expectation
  value of that density is regulated with a suitable number of
  zero-momentum scalar-density insertions on the quark line.  At large
  mass in the hopping parameter expansion, the operator can be
  expressed as a sum of gauge-link loops analogous to those in the
  Boulder discretization, which regulates the construction of $F
  \widetilde F$ through an extended discretization and HYP-smeared
  gauge-links.  In the Boulder case the localization of the gauge paths is
  controlled by the number of smearing steps, whereas localization of
  the L\"uscher operator is controlled by the quark mass.  Of course,
  the chiral properties of the underlying action in that case allows
  an arbitrary choice of scale.

Whatever the definition, the resulting susceptibility is subject
in general to multiplicative and additive corrections at
nonzero lattice spacing \cite{Campostrini:1988cy}:
\begin{equation}
\hat{\chi_t}(a,m_q) = 
M(a,m_q)^2 \, \chi_t(m_q) + A(a,m_q).
\label{eq:renorm}
\end{equation}
An additive renormalization is not required for chiral
actions that use the same operator in the fermion determinant and the
measurement of the topological charge \cite{Durr:2006ky}.  In our case
an additive renormalization is expected.  We assume that in the
continuum limit $M$ approaches one and $A$ approaches zero.  Since
with our actions lattice artifacts appear at ${\cal O}(a^2)$ (up to
logarithms), we expect that the approach to these limits is as $a^2$
\cite{Bernard:2003gq}. With the overlap method one can use the same
Dirac operator for the Monte Carlo evolution and the measurement of
topological charge.  In this case the small instantons and
dislocations that are not seen by the overlap operator, so not
suppressed by a small quark mass, are then also not seen by the
topological charge operator.  In our case the Monte Carlo Dirac
operator and topological charge operators are unrelated, so we might
expect larger lattice artifacts.

\subsection{Predictions from chiral perturbation theory}

Our computed topological susceptibility is a function of the quark
masses and the lattice spacing.  As we have already recalled in
Sec.~\ref{sec:intro}, in chiral perturbation theory the susceptibility
$\chi_t$ depends on the number of light quarks and their masses in
leading order through
\begin{equation}
  \Sigma/\chi_t =  1/m_u + 1/m_d + 1/m_s + \ldots{}.
  \label{eq:chipt}
\end{equation}
where $\Sigma$ is the chiral condensate to this order, $m_u$, $m_d$,
and $m_s$ are the masses of the up, down, and strange quarks, and the
ellipsis represents contributions beyond the cutoff from higher quark
masses and the axial anomaly \cite{Leutwyler:1992yt}.  We see that as
quark masses vanish, the susceptibility must vanish. The rate at which
it vanishes depends on the number of light flavors.

For equal up and down quark masses we may use the
Gell-Mann-Oakes-Renner relation, also from leading order chiral
perturbation theory, to rewrite this expression as
\begin{equation}
  f_\pi^2/(4 \chi_t) = 2/m_\pi^2 + 1/m_{ss}^2 + \ldots{},
 \label{eq:chiral}
\end{equation}
where $m^2_{ss} = 2m_K^2 - m_\pi^2$ is the squared mass of the
fictitious pseudoscalar meson containing two nonannihilating quarks
with masses equal to the strange quark, and in our normalization the
pion decay constant $f_\pi$ is approximately 130 MeV.

With nonchiral lattice fermions, at nonzero lattice spacing one should
instead use a version of chiral perturbation theory appropriate to the
lattice fermion formulation.  In this way some of the lattice
discretization errors can be modeled.  For staggered fermions using
the fourth-root procedure, we use rooted staggered chiral perturbation
theory (\rSXPT) \cite{Aubin:2003mg}.  This theory has a taste
multiplet of sixteen pions.  Among them, only the taste singlet pion
is sensitive to the anomaly and so enters the expression for the
topological susceptibility at leading order.  At tree level the
continuum expression is modified by replacing the pseudoscalar meson
masses by their taste-singlet counterparts \cite{Billeter:2004wx}:
\begin{equation}
  1/\chi_t = (4/f_\pi^2)(2/m_{\pi,I}^2 + 1/m_{ss,I}^2 + 3/m_0^2),
\label{eq:suscept_tb}
\end{equation}
where the subscript $I$ identifies the taste singlet, and through the
term in $m_0$, which is proportional to the $\eta'$ mass at lowest
order, we have introduced an explicit anomaly contribution.  The
standard chiral perturbation theory expression corresponds to
$m_0\to\infty$ (and $a\to 0$); introducing $m_0$ in
Eq.~(\ref{eq:suscept_tb}) is phenomenological because $m_0$ is beyond
the physical cutoff scale of chiral perturbation theory.  At infinite
quark mass we get the quenched topological susceptibility $\chi_{tq}$,
which suggests an alternative phenomenological form
\cite{Durr:2001ty},
\begin{equation}
  1/\chi_t = (4/f_\pi^2)( 2/m_{\pi,I}^2 + 1/m_{ss,I}^2) + 1/\chi_{tq} \, .
\label{eq:suscept_tb2}
\end{equation}

\subsection{Topological charge density operator}

As before \cite{Bernard:2003gq}, we use the topological charge operator
of DeGrand, Hasenfratz, and Kovacs \cite{DeGrand:1997gu} optimized for
SU(3) by Hasenfratz and Nieter \cite{Hasenfratz:1998qk}.  The operator
is constructed from closed ten-link paths of gauge matrices as
follows:
\begin{equation}
  \rho(x) = \sum_{j=1}^2 c^1_j\Tr(1-U_j) + c^2_j[\Tr(1-U_j)]^2 \  .
\end{equation}
Specifically, the operator $U_1$ is constructed from a product along a
path from site $x$ in the sequence of directions
$(x,y,z,-y,-x,t,x,-t,-x,-z)$, summed over rotations and reflections,
and the operator $U_2$, from the directions
$(x,y,z,-x,t,-z,x,-t,-x,-y)$.  Both paths lie inside a $2^4$
hypercube. The coefficients are $c^1_1 = 0.07872507$, $c^1_2 =
0.3173630$, $c^2_1 = -0.1888383$, and $c^2_2 = 0.2854577$.  Hasenfratz
{\it et al.} devised this operator to optimize a match with a
geometric definition of topological charge on a ``typical'' set of
gauge configurations.  The operator also reproduces accurately the
charge of an instanton, provided the instanton radius is larger than
the lattice spacing.  The finer details of the construction of this
operator are unimportant for our purposes, since in the end we take
the continuum limit.

We applied this operator to gauge configurations smoothed by three HYP
steps \cite{Hasenfratz:2001hp}.  From the point of view of the
unsmoothed gauge field, this operation, in effect, enlarges the
footprint of the topological charge density operator by a small
amount.  We have shown in \cite{Bernard:2003gq} that the topological
susceptibility on a coarse lattice ($a \approx 0.12$ fm) is constant
within statistical errors of 8\% for one to four HYP sweeps.

\subsection{Variance reduction method}

\begin{figure}[t]
\begin{center}
\epsfxsize=0.45\textwidth
\epsfbox{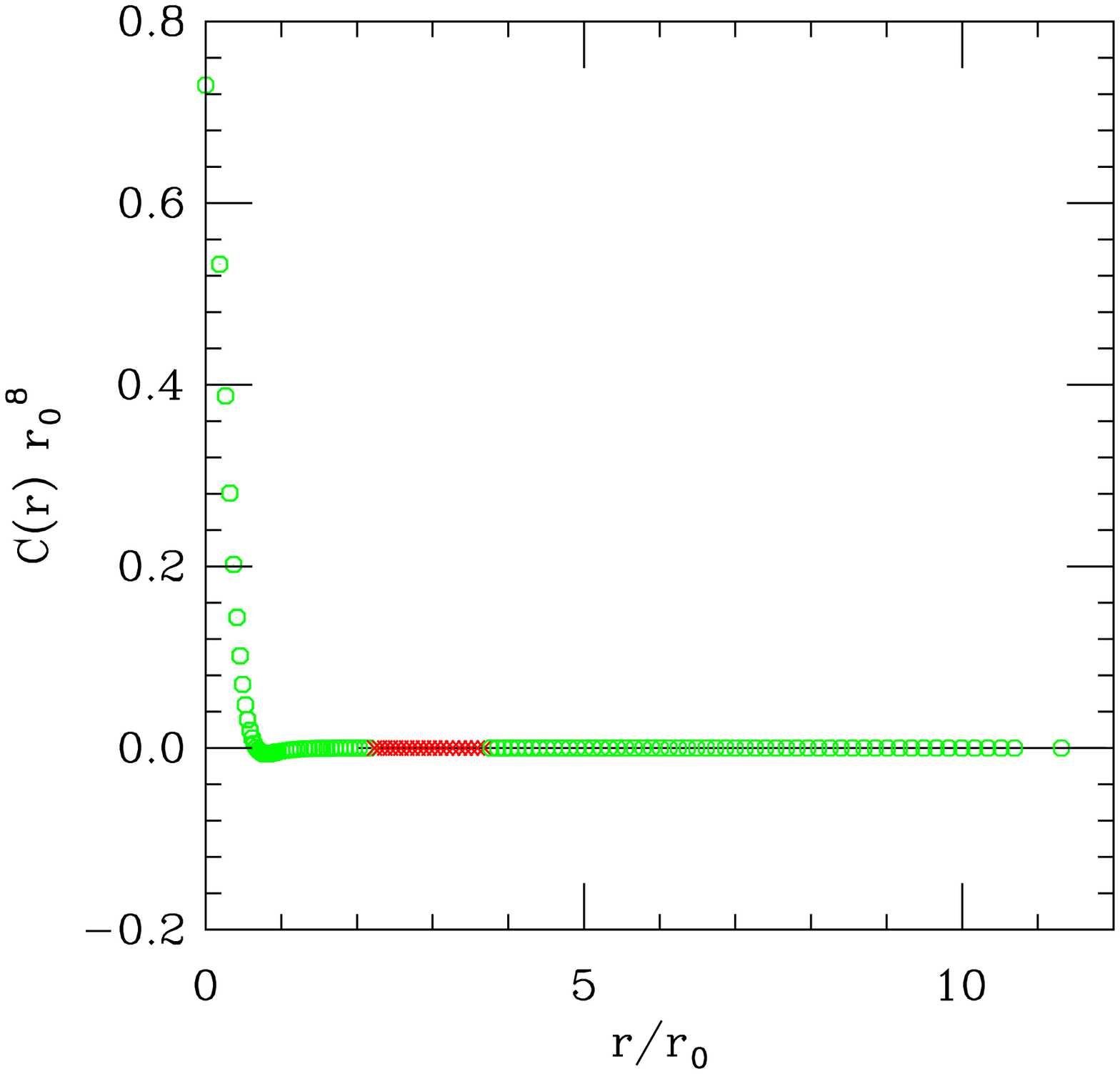}
\hspace{0.02\textwidth}
\epsfxsize=0.43\textwidth
\epsfbox{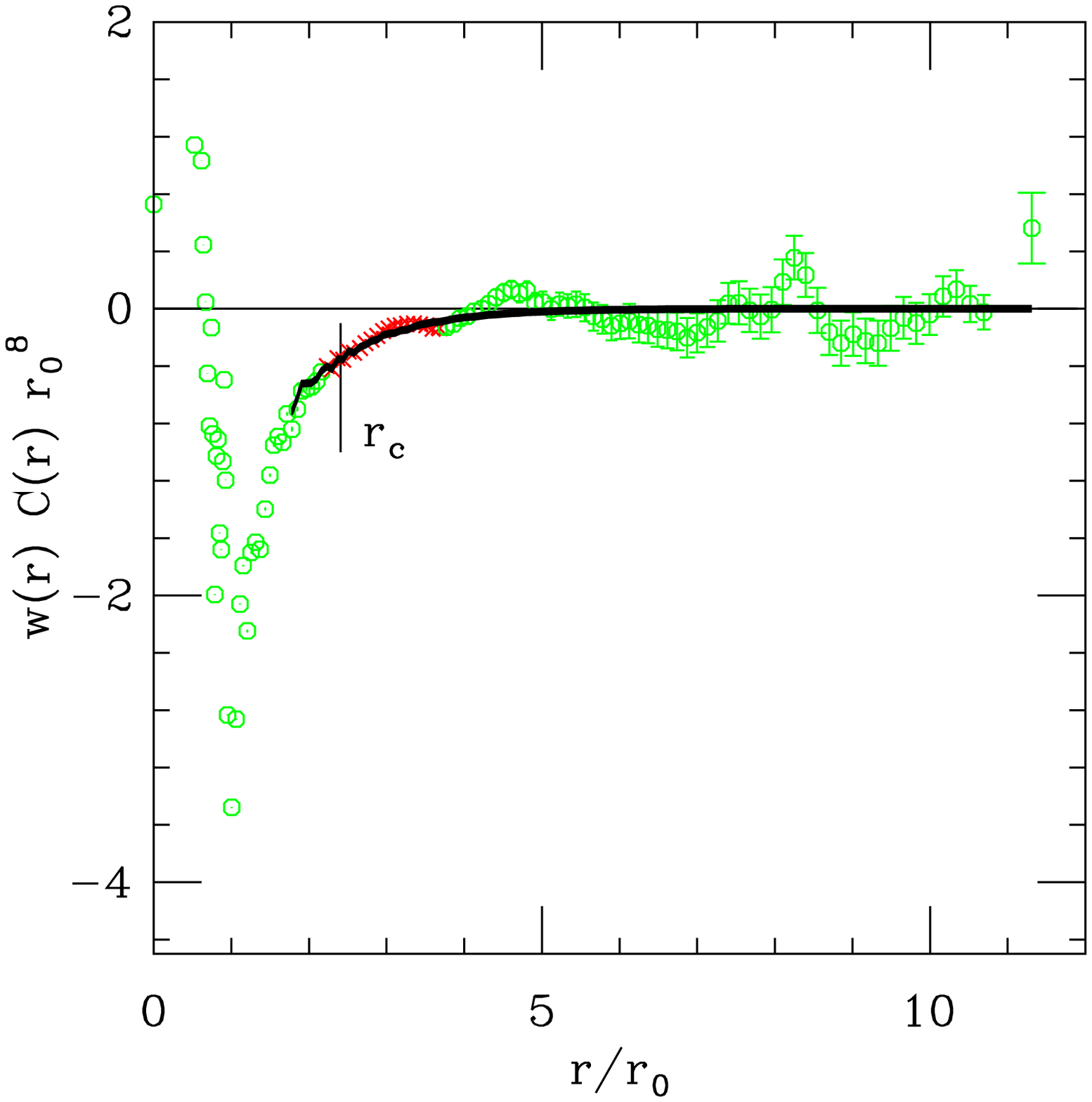}
\end{center}
\caption{Left: Topological charge density correlation function $C(r)$
  {\it vs.}\ separation in units of $r_0$.  Right: Correlation function
  weighted by the volume measure. Errors are statistical and have not
  been corrected for autocorrelations.  The red symbols (crosses)
  indicate the fitted points.  The black curve shows the fit, which we
  use to replace the measured points for $r > r_c$, the cut
  radius. (The lone symbol at the right bins all measurements for
  $r/r_0 > 10.7$). \label{fig:corr_typical}}
\end{figure}

We calculate the topological susceptibility by integrating the
topological charge density correlator in Eq.~(\ref{eq:corrint}) over
the lattice four-volume.  In the left panel of
Fig.~\ref{fig:corr_typical} we show a typical correlator $C(r)$.  It
is expressed in units of the Sommer parameter $r_0 \approx 0.454$ fm
\cite{Sommer:1993ce}.  As expected, it has a positive peak at the
origin next to a negative minimum, and it rises to its asymptotic
limit of zero from below as required by CP symmetry.  To give a better
visual impression of contributions to the susceptibility, in the right
panel of Fig.~\ref{fig:corr_typical} we multiply $C(r)$ by the
statistical weight factor $w(r)$ that counts the number of lattice
points that, by symmetry, have the same four-radius $r$, or, where the
plotted value is binned, have the same range of four-radii.  This is
essentially a discretized version of $r^3 C(r) dr$.  The irregular
binning inherent in the discretized distance $r$ produces the ragged
appearance of the weighted values at small $r$.  On the other hand,
statistical fluctuations produce the ragged appearance at large $r$.
The topological susceptibility in $r_0$ units is simply proportional
to the sum of the weighted values.

In Fig.~\ref{fig:corr_typical}, right, the substantial cancellation of
positive and negative contributions at small $r$ is more evident.  We
also see that the large distance contribution to the susceptibility is
mostly noise.  We have found that it is responsible for the bulk of
the variance in the integral.  This is to be expected.  In a suitably
large subvolume $V_0$ of spacetime, we should be able to determine the
topological susceptibility reasonably well by measuring fluctuations
of the local topological charge $\nu_0$.  Consider putting together
$N$ such volumes to create the total volume $V$.  The overall
topological charge $\nu$ is then obtained as a random walk of local
charges, so its variance grows with $N$.  We can measure the
susceptibility in two ways: (1) average the locally determined
$\langle\nu_0^2\rangle/V_0$ over the $N$ subvolumes or (2) calculate
$\langle\nu^2\rangle/V$ over the full volume.  With the former method
the error in the measured susceptibility decreases as $1/\sqrt{N}$
with increasing $N$ and fixed $V_0$, whereas with the latter
method the error never improves.

In our case the integral of the correlator $C(r)$ replaces the sum
over subvolumes.  But we still need to eliminate noise from
contributions at large $r$.  To do so, several years ago we introduced
a variance-reduction method that fits the large $r$ part of the
correlator to its asymptotic form in Eq.~(\ref{eq:corrasympt}) and then, for
$r > r_c$ for a suitable cutoff $r_c$, replaces the
numerical sum of the correlator with an integral over the fitted
function as follows \cite{Aubin:2004qz}:
\begin{equation}
  \chi_t = \int_{r < r_c}C(r) + \int_{r > r_c} C_{\rm fit}(r) \, ,
\end{equation}
where $r_c$ is chosen inside the fit range.  In the present study we
chose $r_c \approx 1.2$ fm for all ensembles.  In
Fig.~\ref{fig:corr_typical}, right, we illustrate the fit to the large
$r$ part of the correlator and indicate $r_c$.  We continue to use
this method in the present work.

\subsection{Asymptotic fit model}

The topological charge density is a flavor-singlet operator with
quantum numbers $J^{PC} = 0^{-+}$, so the asymptotic behavior of the
correlator is governed by the $\eta$ and $\eta^\prime$ mesons and, for
sufficiently light sea quarks, by multipion states.  That is
\begin{equation}
  C(r) = \langle \rho(x) \rho(0) \rangle \rightarrow A_\eta S(m_\eta,r) + 
  A_{\eta^\prime} S(m_{\eta^\prime},r) + \ldots{},
\label{eq:fitmodel}
\end{equation}
where the $A$'s are overlap constants and $S(m,r)$ is a scalar
propagator with asymptotic form
\begin{equation}
 S(m,r) \approx \left[1 + 3/(8mr)\right] \exp(-mr)/r^{3/2} .
\label{eq:corrasympt}
\end{equation}
The three-pion continuum is the lightest multimeson state in this
correlator.  For our ensembles the $\eta$ meson is always lighter.
Furthermore, the coupling of the topological charge density operator
to multimeson states is Zweig-rule suppressed.  Therefore, we ignore
them in the present analysis.  Since the topological charge density
operator is an SU(3) flavor singlet, it couples to the flavor singlet
component of the $\eta$ and $\eta^\prime$ mesons.  In the usual
representation of singlet-octet mixing \cite{Gilman:1987ax},
\begin{eqnarray}
  \ket{\eta} &=& \cos\theta\ket{\eta_8} + \sin\theta\ket{\eta_0} \nonumber \\
  \ket{\eta^\prime} &=& -\sin\theta\ket{\eta_8} + \cos\theta\ket{\eta_0} ,
\end{eqnarray}
so
\begin{equation}
  A_{\eta}/A_{\eta^\prime} = \tan^2\theta .
\end{equation}

Our statistics are insufficient for determining all the parameters of
the fit function reliably.  Instead, we model the masses of the $\eta$
and $\eta^\prime$ and the ratio $A_\eta/A_{\eta^\prime}$, leaving only
one fitting parameter $A_{\eta^\prime}$, which simply sets the
normalization of the asymptotic form.  We set $m_\eta^2 = 2 m_{\bar s
s}^2/3 + m_\pi^2/3$ for our measured lattice values of $m_{\bar s s}$
and $m_\pi$, and we fix $m_{\eta^\prime} = 958$ MeV (its physical
value) since we have not calculated it for these ensembles.  Finally, we
use a simple chiral model to fix the ratio of couplings
$A_{\eta}/A_{\eta^\prime}$ or equivalently, the singlet-octet mixing
angle as a function of quark masses.

Our model is based on the mass matrix for the flavor-neutral
taste-singlet mesons in lowest order SU(3) chiral perturbation theory
\cite{Aubin:2003mg}
\begin{equation}
  M = \left(
  \begin{array}{ccc}
  M_{UI}^2 + m_0^2 & m_0^2            & m_0^2 \\
  m_0^2            & M_{UI}^2 + m_0^2 & m_0^2 \\
  m_0^2            & m_0^2            & M_{SI}^2 + m_0^2 \\
  \end{array}
   \right),
\end{equation}
where $M_{UI}$ and $M_{SI}$ are masses of unmixed $\bar u u$ ($\bar d
d$) and $\bar s s$ meson states, and $m_0^2$ parameterizes the
anomaly.  The isosinglet eigenvectors are
\begin{eqnarray}
 \ket{\eta} &=& v_u\ket{\bar u u} + v_d\ket{\bar d d} + v_s\ket{\bar s s} \nonumber \\
 \ket{\eta^\prime} &=& v^\prime_u\ket{\bar u u} + v^\prime_d\ket{\bar d d} + 
                v^\prime_s\ket{\bar s s},
\end{eqnarray}
where
\begin{eqnarray}
  v_u &=& v_d = 1/N \nonumber \\
  v_s &=& -(M_{UI}^2 - M_{SI}^2 + m_0^2 + \sqrt{d})/(2m_0^2 N) \nonumber \\
  v^\prime_u &=& v^\prime_d = 1/N^\prime \\
  v_s^\prime &=& -(M_{UI}^2 - M_{SI}^2 + m_0^2 - \sqrt{d})/(2m_0^2 N^\prime) \nonumber \\
    d &=& (M_{SI}^2 - M_{UI}^2)^2 - 2(M_{SI}^2 - M_{UI}^2)m_0^2 + 9m_0^4, \nonumber
\end{eqnarray}
and $N$ and $N^\prime$ normalize the eigenvectors to 1.  Since the
flavor singlet state in this basis is just $(1,1,1)/\sqrt{3}$, we
obtain the ratio
\begin{equation}
  A_{\eta}/A_{\eta^\prime} = \tan^2\theta = 
     (v_u + v_d + v_s)^2/(v^\prime_u + v^\prime_d + v^\prime_s)^2,
\end{equation}
which we apply to the fit model of Eq.~(\ref{eq:fitmodel}).  To
complete the model, we need the value of the anomaly parameter
$m_0^2$.  We set it so that for physical values of $M_{UI}^2$ and
$M_{SI}^2$ ({\it i.e.}, values that give the physical masses $m_\pi$
and $m_{\bar s s} = \sqrt{2m_K^2 - m_\pi^2}$), we get the standard
phenomenological mixing angle $\theta \approx -20$ degrees
\cite{Gilman:1987ax}.  At this ``physical'' point the mixing model
also gives us $m_\eta = 493$ MeV and $m_{\eta^\prime} = 953$ MeV,
reasonably close to their physical values. Then for unphysical masses
we use the lattice values of $M_{UI}^2$ and $M_{SI}^2$ on each
ensemble, always keeping $m_0$ fixed.  This procedure assures that the
$\eta$ decouples as required in the SU(3) flavor limit $m_u = m_d =
m_s$, and it provides a smooth interpolation between that limit and
the physical limit. The taste-singlet masses $M_{UI}^2$ and $M_{SI}^2$
are obtained by adding measured or estimated taste splittings to the
masses of the lightest members of the taste multiplet.  Splittings are
listed in Table~\ref{tab:split} below.

The model is applied to all the dynamical ensembles in this study,
listed in the Appendix \ref{sec:data}. The resulting fit parameters
are listed in Table~\ref{tab:fitparams}. The mixing parameter
$A_{\eta}/A_{\eta^\prime}$ is shown to three digits.  Apart from
systematic errors in the model itself, in principle it inherits a
statistical error from our determination of the taste-singlet masses,
which, in turn depends on the error in the taste splitting.  The last
error, however, is less than 5\%, small enough to have no effect on
the mixing parameter to the number of digits reported.  The remaining
fit parameters do not depend on the taste-singlet masses.
Consequently, statistical errors in the determination of the
taste-singlet masses have negligible effect on results for the
topological susceptibility.

\begin{table}[ht]
\begin{tabular}{lllllrr}
\hline
$10/g^2$ & $m_{ud}/m_s$ & $A_{\eta}/A_{\eta^\prime}$ 
  & $a m_\eta$ & $am_{\eta^\prime}$ & $\chi^2_{\rm raw}$ & $\chi^2/df$\\
\hline
\multicolumn{7}{c}{coarse} \\
  6.85  &  0.05/0.05     & 0.000 & 0.485 & 0.573 &  6.6 &  3.9/11 \\
  6.83  &  0.04/0.05     & 0.010 & 0.470 & 0.578 &  7.6 &  4.5/11 \\
  6.79  &  0.02/0.05     & 0.095 & 0.439 & 0.583 & 17.0 & 10.0/11 \\
  6.76  &  0.01/0.05     & 0.166 & 0.424 & 0.588 &  8.8 &  5.2/11 \\
  6.76  &  0.007/0.05    & 0.194 & 0.417 & 0.584 & 11.1 &  6.5/11 \\
  6.76  &  0.005/0.05    & 0.215 & 0.413 & 0.582 & 11.8 &  6.9/11 \\
\hline                                                            
\multicolumn{7}{c}{fine} \\
  7.18  &  0.031/0.031   & 0.000 & 0.320 & 0.403 & 40.5 & 11.2/19 \\
  7.11  &  0.0124/0.031  & 0.072 & 0.292 & 0.415 & 43.6 & 12.1/19 \\
  7.09  &  0.0062/0.031  & 0.128 & 0.280 & 0.416 & 24.3 &  6.8/19 \\
  7.085 &  0.00465/0.031 & 0.144 & 0.277 & 0.416 & 33.7 &  9.3/19 \\
  7.08  &  0.0031/0.031  & 0.162 & 0.274 & 0.417 & 17.9 &  5.0/19 \\
  7.075 &  0.00155/0.031 & 0.181 & 0.271 & 0.416 & 26.3 &  7.3/19 \\
\hline                                                            
\multicolumn{7}{c}{superfine} \\
  7.48  &  0.0072/0.018  & 0.049 & 0.186 & 0.291 & 52.4 & 16.4/29 \\
  7.475 &  0.0054/0.018  & 0.066 & 0.182 & 0.291 & 29.8 &  9.3/27 \\
  7.47  &  0.0036/0.018  & 0.087 & 0.178 & 0.291 & 30.4 &  9.5/27 \\
  7.465 &  0.0025/0.018  & 0.101 & 0.175 & 0.291 & 26.5 &  8.2/26 \\
  7.46  &  0.0018/0.018  & 0.110 & 0.174 & 0.292 & 35.2 & 11.0/26 \\
\hline                                                            
\multicolumn{7}{c}{ultrafine} \\
  7.81  &  0.0028/0.014  & 0.097 & 0.136 & 0.216 & 29.1 & 14.6/30 \\
\end{tabular}
\caption{Parameters used in asymptotic fits to the $(2+1)$-flavor
  topological charge density correlator.  The raw $\chi^2$
  is uncorrected for autocorrelations.  The last column includes the
  correction as explained in Sec.~\ref{subsec:autocorr}.}
  \label{tab:fitparams}
\end{table}

\subsection{Asymptotic fit model for the quenched ensembles}

For the three quenched ensembles we use the same methodology, except
that the fit model has only one mass.  We fix it to the mass of the
$J^{PC} = 0^{-+}$ ground state lattice glueball from Chen {\it et al.}
\cite{Chen:2005mg}, namely 2560 MeV.  The parameters are listed in
Table \ref{tab:fitparams2}.  We chose $r_c$ for the quenched ensembles
to match our choice for the dynamical ensembles at the same lattice
spacing.  Since the quenched correlators die so quickly at large $r$,
the contribution to the susceptibility for $r > r_c$ is negligible,
and the asymptotic model has no effect on the result.

\begin{table}[ht]
\begin{tabular}{llr}
\hline
$10/g^2$ & $a m_G$ & $\chi^2/df$ \\
\hline
  8.00 &   1.55  &  16.0/12 \\
  8.40 &   1.11  &   9.8/11 \\
  8.80 &   0.816 &  10.0/13 \\
\hline
\end{tabular}
\caption{Parameters used in asymptotic fits to the quenched
  topological charge density correlator and resulting values of
  $\chi^2/df$.}
\label{tab:fitparams2}
\end{table}

\section{Results}

\label{sec:results}

We smooth the lattices with three HYP smoothing steps
\cite{Hasenfratz:2001hp} and measure the topological charge density
with the Boulder operator at each space-time point.  We then construct
the point-to-point correlator $C(r)$ for every pair of points in the
space-time volume.  For $r/a < 5$ we keep values for every
displacement, and for larger $r$ we bin data over small intervals in
$r$.  The resulting data is then fit to Eq.~(\ref{eq:fitmodel}) over a
range $[r_{\rm min},r_{\rm max}]$.  We replace the raw data with the
fit model for $r > r_c$. The fit range is chosen to give an
acceptable $\chi^2/df$ (corrected for autocorrelations) and to vary
smoothly as a function of sea quark mass and lattice spacing.

\subsection{Monte Carlo time histories and autocorrelations}
\label{subsec:autocorr}

\begin{figure}[t]
\begin{center}
\epsfxsize=0.8\textwidth
\epsfbox{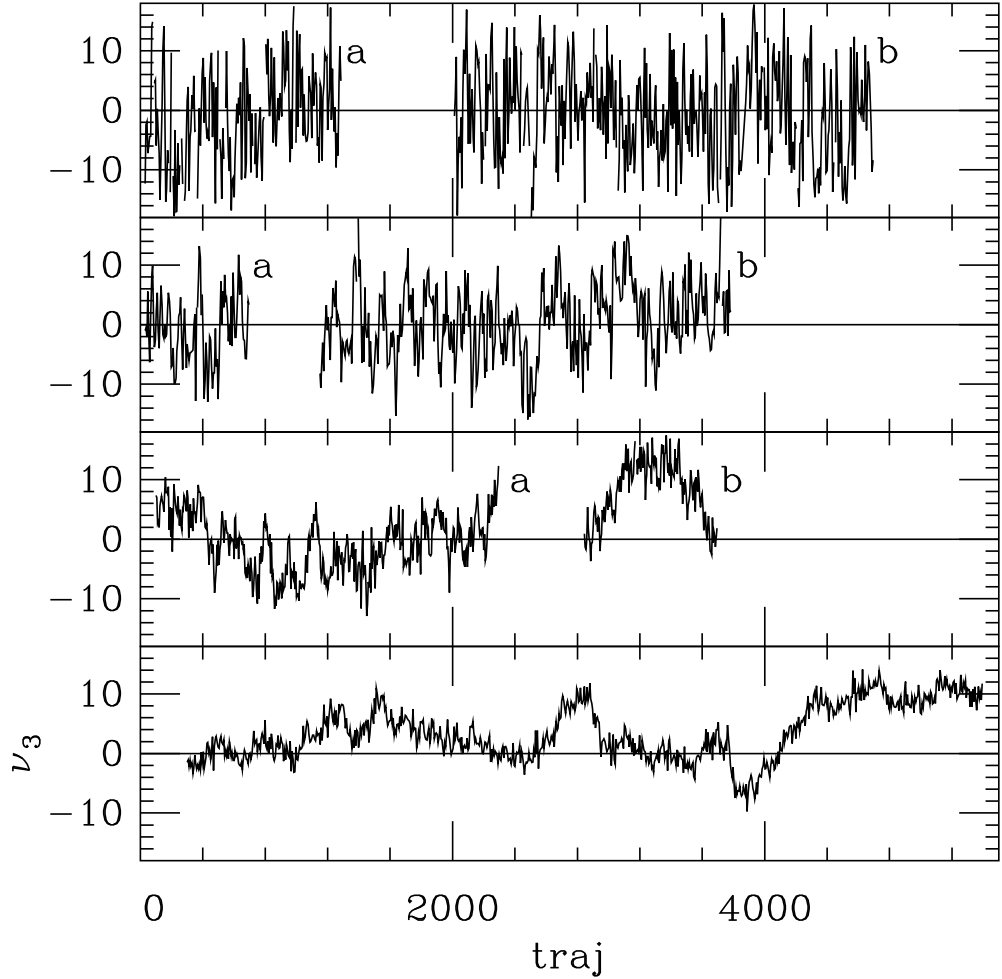}
\end{center}
\caption{Total topological charge after three HYP sweeps as a function
  of simulation time for four lattice spacings and fixed sea quark
  masses with ratio $m_{ud}/m_s = 0.2$.  Sections marked ``a'' and ``b''
  come from different Markov chains.  From top to bottom, $a = 0.12$, 0.09,
  0.06, and 0.045 fm.}
\label{fig:qtot_history}
\end{figure}

\begin{figure}[t]
\begin{center}
\epsfxsize=0.8\textwidth
\epsfbox{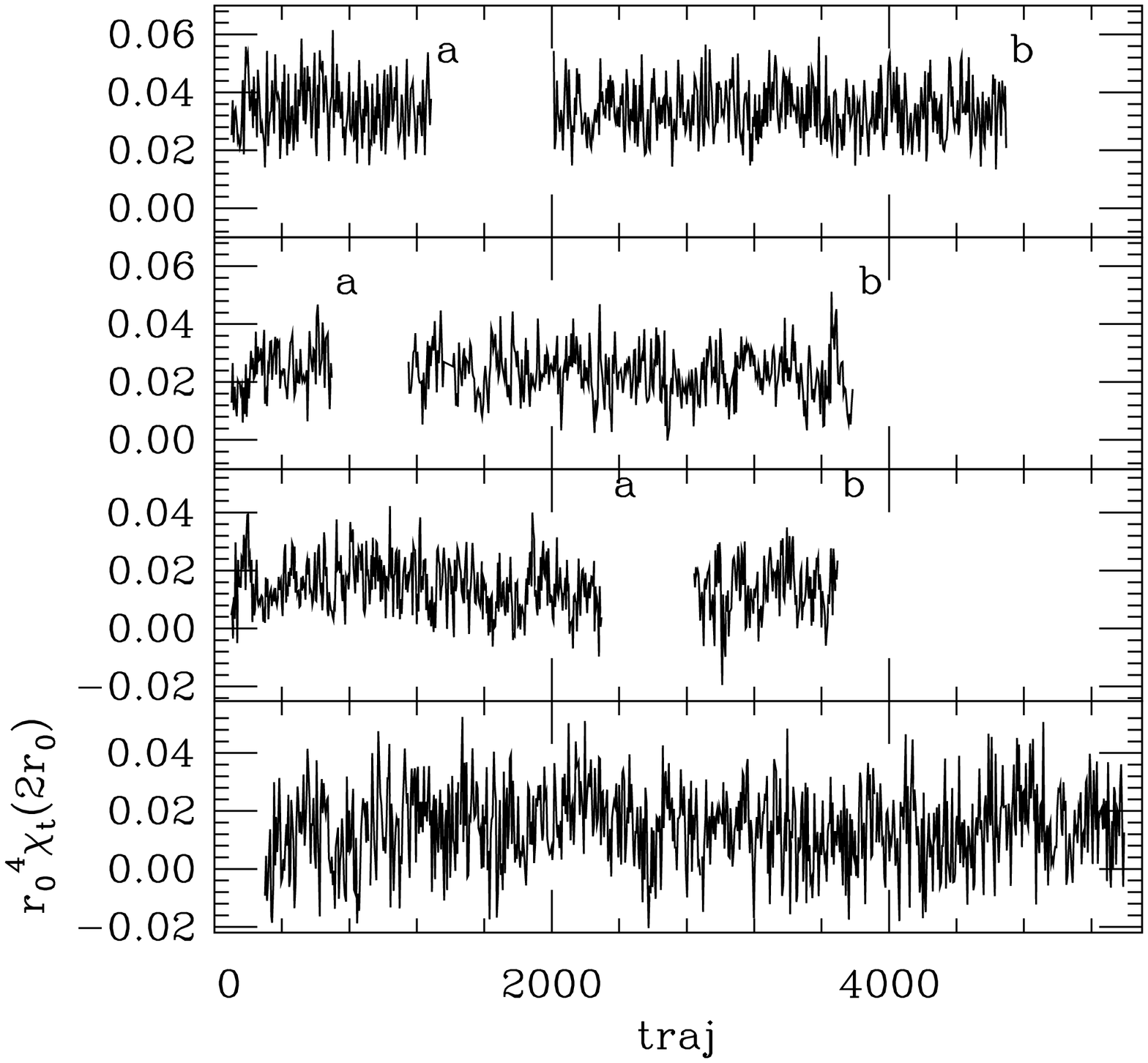}
\end{center}
\caption{Contribution to the topological susceptibility for $r < 2r_0$
as a function of simulation time for the ensembles of
Fig.~\protect\ref{fig:qtot_history}. \label{fig:chi_r0_history}}
\end{figure}

To determine the confidence level of our fits and errors in the fit
parameters, we must first analyze autocorrelations in Monte Carlo
time.  With our action and molecular dynamics algorithm, the total
topological charge is moderately persistent in Monte Carlo time.  In
Fig.~\ref{fig:qtot_history}, we show the time histories for a range of
lattice spacings for $m_{ud} = 0.2 m_s$ ensembles.  As we have noted,
however, the topological susceptibility is a local observable.  We can
get a graphical sense of the autocorrelation affecting the
susceptibility by considering the time history of the integral of the
correlator
\begin{equation}
  \chi_t(r) = \int_0^r  C(r^\prime) \, 2\pi^2 (r^\prime)^3 dr^\prime.
\label{eq:chi_r}
\end{equation}
In Fig.~\ref{fig:chi_r0_history} we show the time history of this
variable for the case $r = 2 r_0$ for the same set of ensembles.
Clearly the fluctuations in this quantity decorrelate much more
rapidly than those of the total topological charge.

We estimate the autocorrelation correction, \ie, the amount by which
the naive (uncorrelated) variance should be increased to compensate
for autocorrelations.  For this purpose we consider the integral of
the correlator over the proposed fit range
\begin{equation}
   \int_{r_{\rm min}}^{r_{\rm max}}C(r^\prime)2\pi^2 (r^\prime)^3 dr^\prime.
\end{equation}
We block the data in Monte Carlo time and calculate the variance of
the mean as a function of block size, extrapolating to infinite block
size.  The ratio of the extrapolated variance to the naive variance is
the correction factor.  We also sum the autocorrelation coefficients
to obtain another estimate of the correction factor.  These
determinations fluctuate as a function of sea quark mass.  We averaged
them at fixed lattice spacing to obtain the correction factors shown
in Table~\ref{tab:autocorr}.  We should emphasize that the
determination of autocorrelation corrections is notoriously difficult.
To develop more confidence in these estimates, we should have
considerably longer time histories.

\begin{table}[ht]
\begin{tabular}{ll}
\hline
spacing   & correction \\
\hline
coarse    & 1.7 \\
fine      & 3.6 \\
superfine & 3.2 \\
ultrafine & 2.0 \\
\hline
\end{tabular}
\caption{Autocorrelation correction factors for the various categories
of lattice spacings in this study.  The factor multiplies the naive
variance.}  \label{tab:autocorr}
\end{table}

Our fits to the data take into account correlations in $r$ as well.
For all ensembles, measurements are taken every six or sometimes every
five molecular dynamics time units. We do not bin data in Monte Carlo
time before constructing the covariance matrix in $r$ and minimizing
the correlated $\chi^2$ \cite{Toussaint:2008ke}.  Uncorrected errors
are derived from a jackknife analysis.  Thus the resulting $\chi^2$,
based on the naive covariance, must be reduced by the factor in Table
\ref{tab:autocorr} before estimating the confidence level.
Furthermore, the naive single-elimination jackknife errors in the fit
parameters must be increased by the square root of this factor.  We
use the same factor to adjust the error in the contribution from the
raw data for $r < r_c$.

\subsection{Topological charge density correlator}

\begin{figure}[t]
\begin{center}
\epsfxsize=0.7\textwidth
\epsfbox{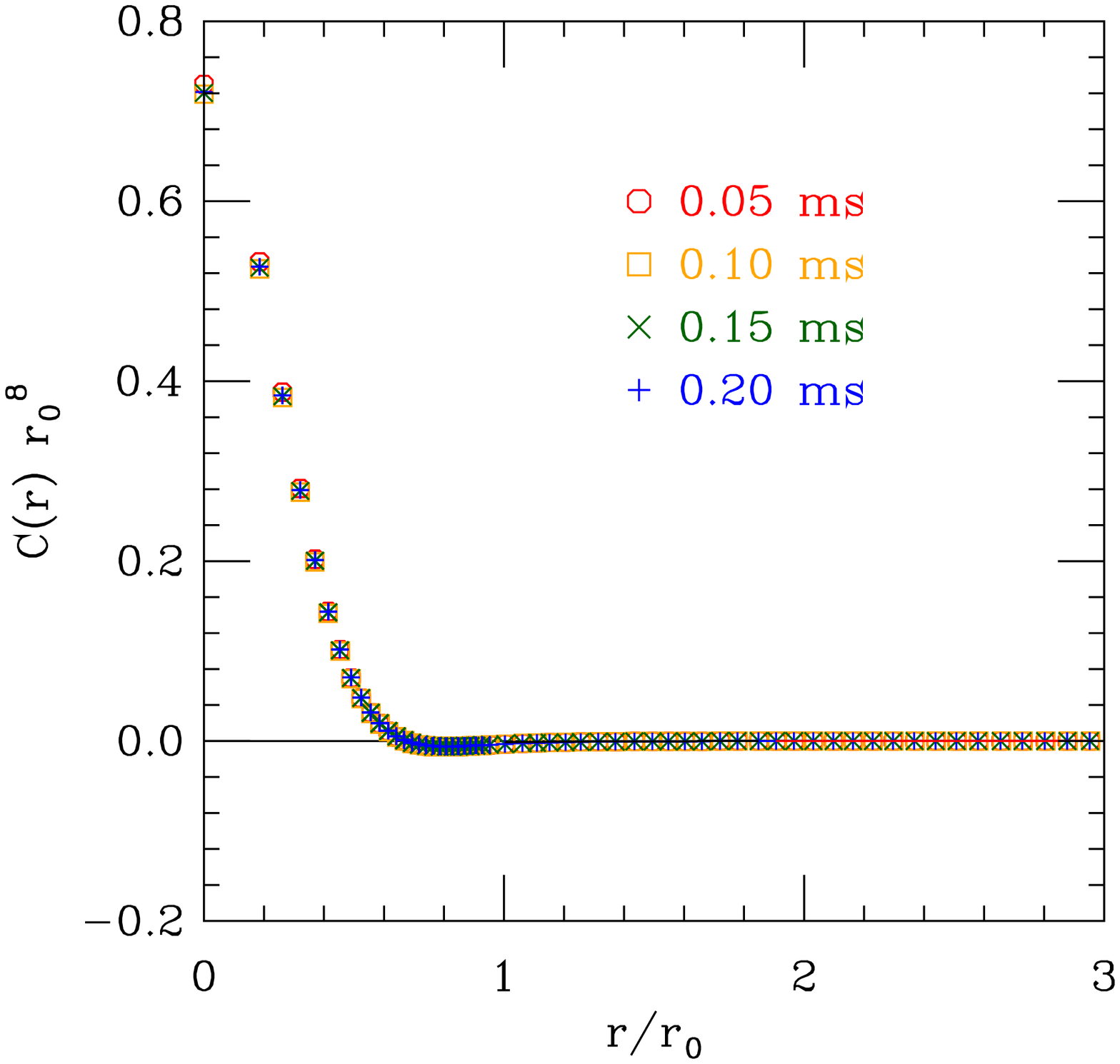}
\\
\vspace*{10mm}
\epsfxsize=0.7\textwidth
\epsfbox{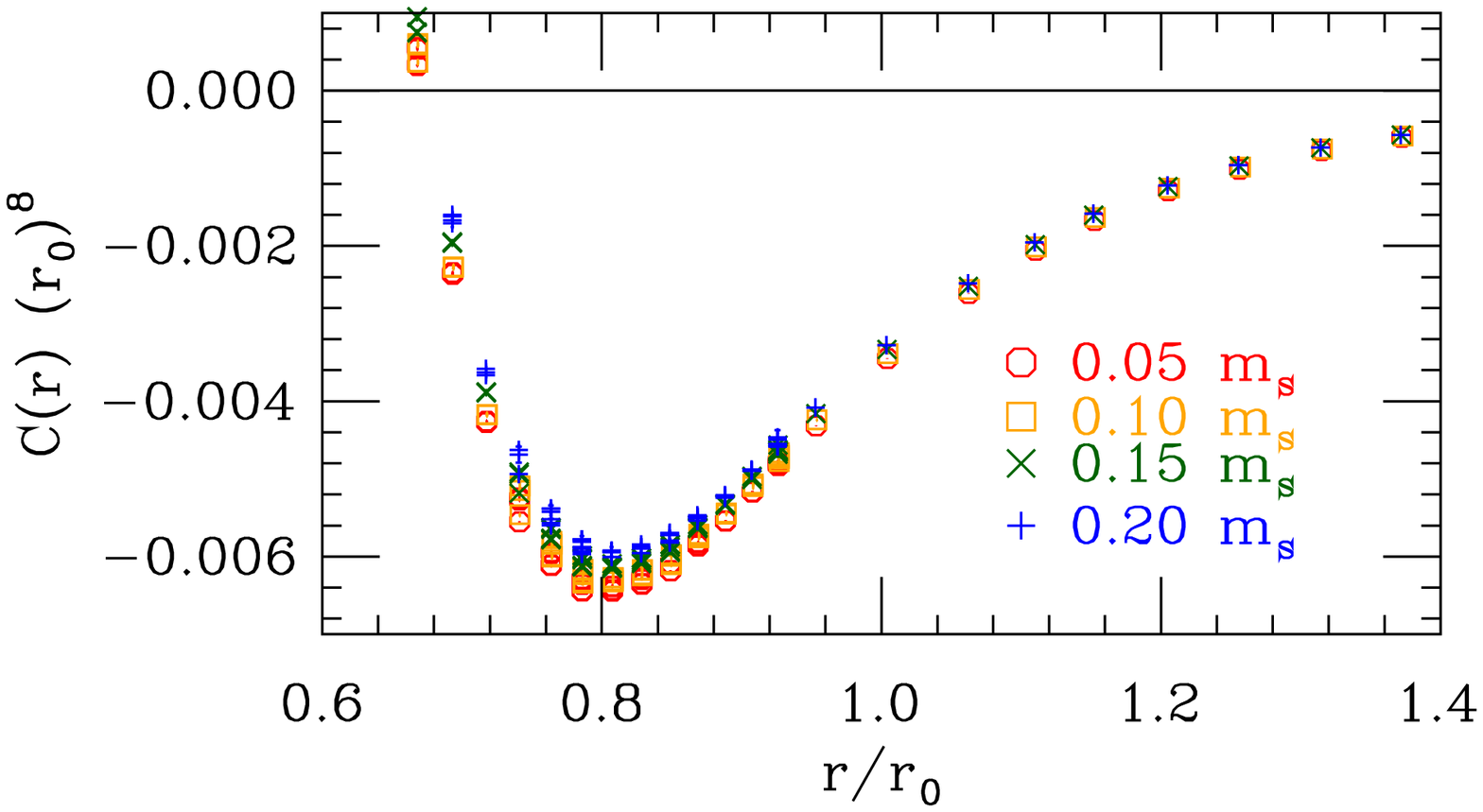}
\end{center}
\caption{Topological charge density correlator {\it vs.}\ $r$ in units of
$r_0$ for a set of fine lattice ensembles ($a \approx 0.09$ fm) with
varying light sea quark masses $m_{ud}$.  Upper: overview.  Lower:
detail. \label{fig:corr_vs_r_mud} }
\end{figure}

We expect the topological susceptibility to decrease with decreasing
light sea quark mass.  It is interesting to see how the topological
charge density correlator itself varies with the light sea quark mass
at fixed lattice spacing.  In Fig.~\ref{fig:corr_vs_r_mud} we examine
this dependence for a series of fine lattice ensembles ($a \approx
0.09$ fm) for which we have results for four ratios of the light to
strange quark mass, $0.05$, $0.1$, $0.15$, and $0.2$,
corresponding to the range 0.601 to 1.074 in $m_{\pi
I}^2r_0^2$.  In the upper panel any variation with light quark mass
is evidently much smaller than the plot symbol size.  In fact the
short distance part of the correlator shows very little sea quark mass
dependence.  In the lower panel we enlarge the region around the
minimum where a small variation is now apparent.  In this region light
meson states begin to dominate the correlator of the gluonic
operators.  As the light quark mass decreases, the minimum drops, thus
giving a larger negative contribution to the integral.  This effect
leads to the suppression of the susceptibility. According to the
model, the correlator should also decay more slowly at large $r$, but
this effect is too subtle to be visible with our statistics.

\begin{figure}[t]
\begin{center}
\epsfxsize=0.55\textwidth
\epsfbox{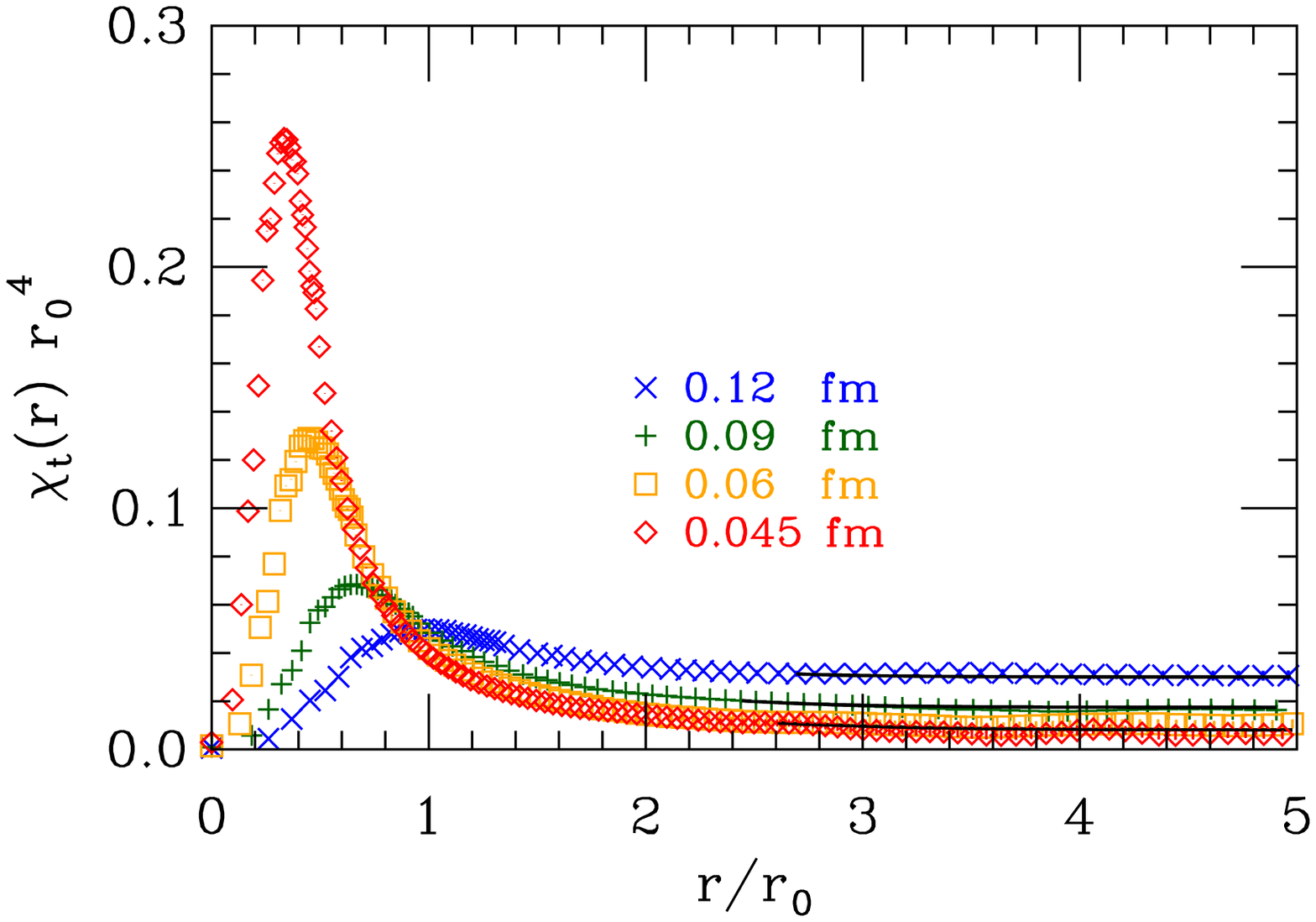} 
\\
\vspace*{10mm}
\epsfxsize=0.65\textwidth
\epsfbox{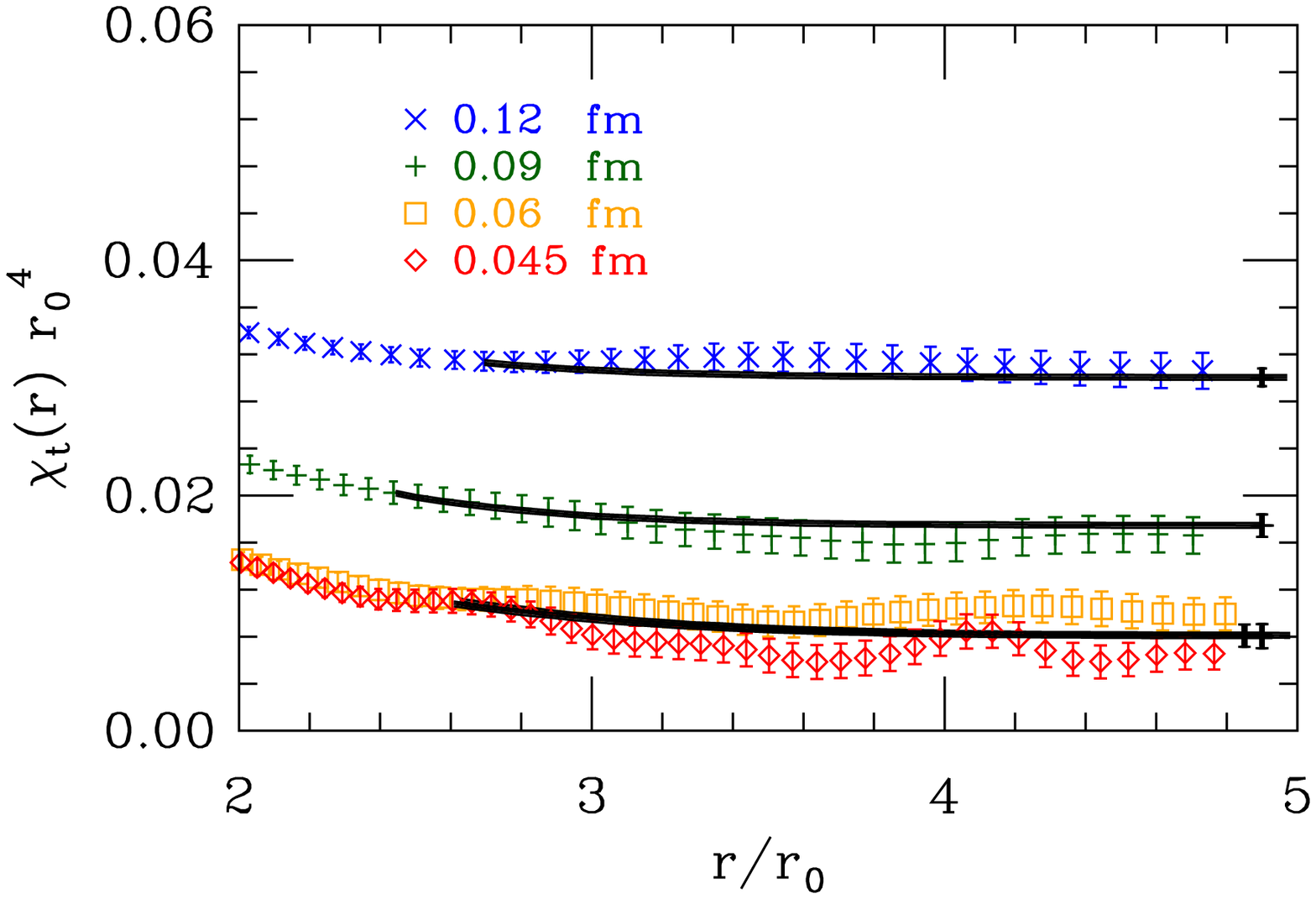} \\
\end{center}
\caption{Upper panel: integrated topological density correlator
  $\chi_t(r)r_0^4$ {\it vs.}\ $r/r_0$ at fixed light quark mass $m_{ud}
  = 0.2 m_s$ for the lattice spacings indicated. Lower panel: detail
  of the asymptotic behavior. The full topological susceptibility is
  the value at the largest $r$. The plotted points give the result
  from the raw data without variance reduction.  Errors include the
  adjustment for autocorrelations listed in Table~\ref{tab:autocorr}.
  The solid black curves show the central value of the integrated
  contribution with the fit values replacing the raw data for $r >
  r_c$. (Values of $r_c$ and fit ranges are given in Table
  \protect\ref{tab:results}.)  The fit curves for $a = 0.06$ and 0.045
  fm are, accidentally, nearly coincident.  Statistical errors on the
  solid lines are shown on the right edge of the right panel. They
  have also been corrected for autocorrelations.  The fit error for
  the smallest lattice spacing has the largest error bar. The
  improvement in variance is evident.} \label{fig:cumcorr_vs_r_a}
\end{figure}

We next examine the lattice spacing dependence of the correlator at
fixed light quark mass ratio.  Comparing the local correlators $C(r)$
obtained on ensembles at different lattice spacing is complicated
because sampling is naturally done on a lattice scale.  Rather than
rebinning the data to a common physical scale, we compute the partial
integral $\chi_t(r)$ of Eq.~(\ref{eq:chi_r}) and plot it in physical
($r_0$) units in Fig.~\ref{fig:cumcorr_vs_r_a}.  As $r$ increases from
the origin, we see a peak at short distance coming from the regulated
contact term followed by a decrease coming from the negative
correlator.  The onset and width of the peak is determined by the
effective radius of the topological charge density operator, which is
fixed in lattice units.  Thus as the lattice spacing decreases, the
expected negative $1/r^8$ singularity in the correlator is exposed,
and the peak increases in height and decreases in width.

At large $r$ the data approach the asymptotic value of the full
susceptibility.  The figure shows both the integrated raw data and the
integral with the fit values for $r > r_c$ replacing the raw
data.  The lower panel enlarges the asymptotic region to show the
variance reduction achieved by the fit.  The result also shows a
plausible convergence of the asymptotic value in the continuum limit.

\begin{table}[ht]
\begin{tabular}{llrl}
\hline
spacing   & $a$ (fm) & $r_0/a$ & $\sigma_{\rm corr}$ \\
\hline                   
coarse    & 0.12     & 3.82    & $1.8\times 10^{-4}$ \\
fine      & 0.09     & 5.40    & $2.5\times 10^{-4}$ \\
superfine & 0.06     & 7.73    & $3.3\times 10^{-4}$ \\
ultrafine & 0.045    & 10.39   & $4.4\times 10^{-4}$ \\
\hline
\end{tabular}
\caption{Error $\sigma_{\rm corr}$ in $\chi_t(r_0)$, the
short-distance contribution to the topological susceptibility, at sea
quark mass $m_{ud} = 0.2 m_s$ for various lattice spacings.  The error
is adjusted to the same sample size, autocorrelation, and lattice
volume.}  \label{tab:origin}
\end{table}

Now we point out a practical issue relevant to future extensions of
this work, namely, whether the topological susceptibility, defined by
integrating the correlator of the regulated topological charge density
operator, has a feasibly accessible continuum limit.  This will be the
case if the variance in the integral of the correlator for fixed
physical volume and statistical sample size does not diverge as the
lattice spacing decreases.  We examine $\chi_t(r)$ at a fixed physical
distance $r$ as the lattice spacing decreases.  For $r < r_0/2$ we
find that the variance actually decreases for $a \in [0.045, 0.12]$
fm.  But for such a small range in $r$, the behavior of the integrated
correlator is strongly influenced by the size of the topological
charge density operator.  The larger radius $r = r_0$ is safely
outside the width of the operator and in a region where, for $a \in
[0.045, 0.12]$ fm, the integrated correlator $\chi_t(r)$ is well past
the peak, as we can see from Fig.~\ref{fig:cumcorr_vs_r_a}.  We show
the error in $\chi_t(r_0)$ as a function of lattice spacing in Table
\ref{tab:origin}.  This statistical error is adjusted for
autocorrelations, sample size (factor of $\sqrt{N/N_0}$), and lattice
volume (factor of $\sqrt{V/V_0}$) for $N_0 = 500$ and $V_0 = 100$
fm$^4$.  We see that the adjusted error grows approximately as $1/a$
over this range.  This trend suggests that it will be increasingly
expensive to push to smaller lattice spacing with our scheme.
However, the continuum limit is nonetheless finite, and our results
demonstrate that the method gives reasonable errors over the range of
lattice spacings considered.

\begin{table}
\begin{tabular}{llrl}
\hline
spacing   & $a$ (fm) & $r_0^2 \Delta M^2$\\
\hline               
coarse    & 0.12     & 1.136 \\
fine      & 0.09     & 0.437 \\
superfine & 0.06     & 0.143 \\
ultrafine & 0.045    & 0.087 \\
\hline
\end{tabular}
\caption{\label{tab:split}Mass splittings (difference in squared
  masses) between Goldstone and taste singlet pions}
\end{table}

\begin{figure}[t]
\begin{center}
\epsfxsize=0.7\textwidth
\epsfbox{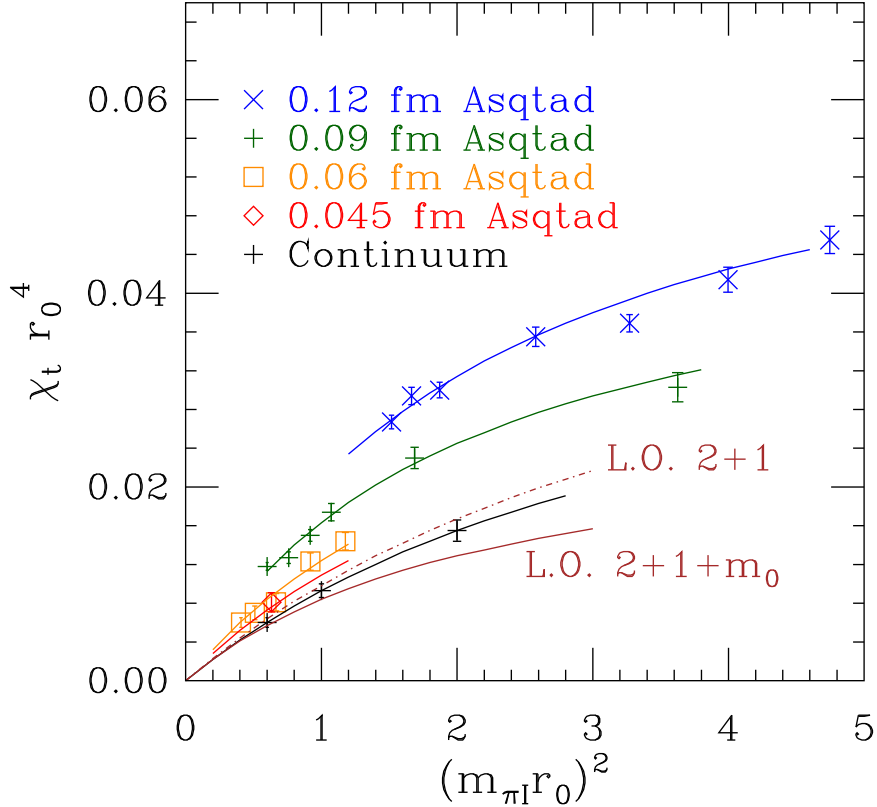} \\
\end{center}
\caption{Topological susceptibility {\it vs.}\ the squared
  taste-singlet pion mass in units of the Sommer parameter $r_0
  \approx 0.454$ fm \protect\cite{Sommer:1993ce}. The brown curve
  labeled ``L.O. $2+1+m_0$ shows the prediction of tree-level continuum chiral
  perturbation theory from Eq.~(\protect\ref{eq:suscept_tb2}) with
  $f_\pi = 130$ MeV, and the dashed brown line labeled
  ``L.O. $2+1$'' shows the same prediction without the last term of
  Eq.~(\protect\ref{eq:suscept_tb2}).  The remaining curves are fits
  to the model of Eq.~(\protect\ref{eq:chitfit}). The solid black line
  is the central value of the continuum extrapolation of that model
  and three representative points on the curve indicate the one sigma
  error.  \label{fig:results}}
\end{figure}

\begin{table}
  \begin{tabular}{llrrrrrr}
\hline
$10/g^2$ & $m_{ud}/m_s$ & range ($a$) & $r_c/a$ & $r_0^2m_{\pi I}^2$ 
& $(\chi_{t<})r_0^4$ & $(\chi_{t>})r_0^4$ & $\chi_t r_0^4$ \\
\hline
\multicolumn{8}{c}{coarse} \\
6.85  &  0.05/0.05     & $[ 8.0, 12]$ & 10 & 4.746  & 0.0461(14) & $-0.0006(2)$ & 0.0455(14) \\
6.83  &  0.04/0.05     & $[ 8.0, 12]$ & 10 & 3.997  & 0.0422(13) & $-0.0008(2)$ & 0.0414(13) \\
6.79  &  0.02/0.05     & $[ 8.0, 12]$ & 10 & 2.580  & 0.0364(10) & $-0.0009(1)$ & 0.0355(10) \\
6.76  &  0.01/0.05     & $[ 8.0, 12]$ & 10 & 1.872  & 0.0315(08) & $-0.0015(1)$ & 0.0300(08) \\
6.76  &  0.007/0.05    & $[ 8.0, 12]$ & 10 & 1.665  & 0.0309(09) & $-0.0015(1)$ & 0.0294(09) \\
6.76  &  0.005/0.05    & $[ 8.0, 12]$ & 10 & 1.517  & 0.0289(07) & $-0.0021(1)$ & 0.0267(07) \\
8.00  &  quenched      & $[ 6.0, 10]$ & 10 & $-$\ \ & 0.0733(08) & $ 0.0000(0)$ & 0.0598(10) \\
\hline
\multicolumn{8}{c}{fine} \\
7.18  &  0.031/0.031   & $[10.0, 18]$ & 13 & 3.626  & 0.0321(13) & $-0.0018(6)$ & 0.0303(15) \\     
7.11  &  0.0124/0.031  & $[10.0, 18]$ & 13 & 1.688  & 0.0247(09) & $-0.0017(4)$ & 0.0230(11) \\     
7.09  &  0.0062/0.031  & $[10.0, 18]$ & 13 & 1.074  & 0.0206(09) & $-0.0031(4)$ & 0.0174(09) \\     
7.085 &  0.00465/0.031 & $[10.0, 18]$ & 13 & 0.918  & 0.0188(06) & $-0.0038(2)$ & 0.0150(06) \\    
7.08  &  0.0031/0.031  & $[11.0, 19]$ & 13 & 0.760  & 0.0170(06) & $-0.0044(4)$ & 0.0127(06) \\
7.075 &  0.00155/0.031 & $[12.0, 18]$ & 13 & 0.601  & 0.0166(02) & $-0.0047(2)$ & 0.0118(04) \\
8.40  &  quenched      & $[ 8.0, 12]$ & 10 & $-$\ \ & 0.0722(07) & $-0.0000(0)$ & 0.0593(10) \\
\hline
\multicolumn{8}{c}{superfine} \\
7.48  &  0.0072/0.018  & $[12.0, 25]$ & 20 & 1.177  & 0.0167(09) & $-0.0023(2)$ & 0.0144(09) \\
7.475 &  0.0054/0.018  & $[12.5, 25]$ & 20 & 0.920  & 0.0148(09) & $-0.0025(2)$ & 0.0123(09) \\
7.47  &  0.0036/0.018  & $[12.5, 25]$ & 20 & 0.666  & 0.0113(09) & $-0.0032(2)$ & 0.0081(09) \\
7.465 &  0.0025/0.018  & $[13.0, 25]$ & 20 & 0.510  & 0.0107(07) & $-0.0037(2)$ & 0.0070(07) \\
7.46  &  0.0018/0.018  & $[13.0, 25]$ & 20 & 0.408  & 0.0100(05) & $-0.0040(2)$ & 0.0060(05) \\
8.80  &  quenched      & $[15.0, 21]$ & 15 & $-$\ \ & 0.0680(06) & $-0.0001(2)$ & 0.0561(12) \\
\hline
\multicolumn{8}{c}{ultrafine} \\
7.81  &  0.0028/0.014  & $[16.0, 32]$ & 27 & 0.634  & 0.0111(10) & $-0.0030(1)$ & 0.0080(10) \\
\end{tabular}
\caption{Fit ranges and cut radius in lattice units and
  results for the topological susceptibility.  Also shown are the
  computed or estimated taste-singlet squared pion masses in $r_0$
  units and the contributions to the total topological susceptibility
  for distances less than ($\chi_{t<}$) and greater ($\chi_{t>}$) than
  the cut radius.}  \label{tab:results}
\end{table}

\subsection{Topological susceptibility}

Our results are summarized in Table~\ref{tab:results} and
Fig.~\ref{fig:results}.  Since chiral perturbation theory predicts the
behavior as a function of the mass of the taste-singlet pion, we also
list estimates of that mass.  Unlike the Goldstone pion mass, the mass
of the taste singlet is not measured directly on all of our ensembles.
However, to a good approximation, splittings of the squared masses of
the pion taste multiplet are known to be independent of the light
quark mass at fixed lattice spacing \cite{Bernard:2001av}.  So if the
splitting is measured for one light quark mass for a given lattice
spacing, the taste-singlet pion mass can be reconstructed from the
Goldstone pion mass for other light quark masses at the same spacing.
Table~\ref{tab:split} lists the splittings for the categories of
lattice spacings in this study.  They were used to obtain the values
in Table~\ref{tab:results}.  The largest error in the estimated
splittings is less than 5\%, which bounds the error in the abscissa of
the plot.  We have chosen $r_c$ to lie within the fit
range. We have found that within this range our results vary by less
than one standard deviation.

\subsection{Continuum extrapolation}

To model a continuum extrapolation, we fit our data to the following
form:
%
%
\begin{equation}
  1/\chi_t = c_0 + c_1 (a/r_0)^2 + 
   [c_2 + c_3 (a/r_0)^2 + c_4(a/r_0)^4]/(m_{\pi,I}r_0)^2 \, .
\label{eq:chitfit}
\end{equation}
This model assumes that lattice artifacts scale as $a^2$.  The fit
yields $\chi^2/df = 8.8/13$.  In Fig.~\ref{fig:results}
the resulting fit curves are shown, and three representative points in
the continuum extrapolation are also plotted.  Also plotted is the
prediction of Eq.~(\ref{eq:suscept_tb2}) using $f_\pi r_0 = 130 \times
0.454$ MeV-fm with and without our continuum-extrapolated asymptotic
quenched topological susceptibility $\chi_t r_0^4 = 0.0523(29)$.  From
the fit itself we obtain $f_\pi = 132(6)$ MeV, which is better than
expected for tree-level chiral perturbation theory.

\section{Conclusions}
\label{sec:conclusions}

We have presented an extensive study of the topological susceptibility
on 18 $(2+1)$-flavor asqtad lattice ensembles and three quenched
lattice ensembles.  The susceptibility is defined as the integral of
the correlator of the topological charge density.  The topological
charge density is constructed from a discretized version of $F
\widetilde F$ with smearing to help regulate ultraviolet fluctuations.
To reduce the variance from large distances, we replace the measured
values of the correlator at large $r$ by a fit model that builds in
the expected spectral contribution.

Our method for determining the topological susceptibility through an
integral of the topological charge density correlator avoids
singularities at zero separation by smearing the charge density
operator over a fixed local set of lattice sites.  A study of the
variance in the small-distance contribution suggests that as the
lattice spacing is decreased the variance grows.  At our level of
statistics and for the range of lattice spacings we consider in this
study, this growth is manageable.

Over the range of lattice spacings and masses in this study, within
statistical errors, we find good agreement with tree-level staggered
chiral perturbation theory and in the continuum limit with tree-level
continuum chiral perturbation theory, in both cases with the expected
number of flavors.  This agreement supports the assertion that the
fourth-root procedure for staggered fermions results in the correct
number of sea quark species in the continuum limit.

\section*{Acknowledgments}

This work was supported by the U.S. Department of Energy under grant
numbers
DE-FC02-06ER-41439, 
DE-FC02-06ER-41443, 
DE-FC02-06ER-41446, 
DE-FC06-01ER-41437, 
DE-FG02-04ER-41298, 
DE-FG02-91ER-40628, 
and
DE-FG02-91ER-40661  
and by the U.S. National Science Foundation under grant numbers
OCI08-32315, 
PHY05-55234, 
PHY05-55235, 
PHY05-55243, 
PHY05-55397, 
PHY07-03296, 
PHY07-04171, 
PHY07-57035, 
PHY07-57333, 
PHY09-03536, 
and          %
PHY09-03571. 
An allocation of computer time from the Center for High
Performance Computing at the University of Utah is gratefully
acknowledged. Computation for this research was supported in part by
the U.S. National Science Foundation through TeraGrid resources
provided by the Texas Advanced Computing Center (TACC), the National
Institute for Computational Sciences (NICS), the National Center for
Supercomputing Applications (NCSA), and the Pittsburgh Supercomputing
Center (PSC) under grant number TG-MCA93S002. Computation for this
work was also carried out on the Fermilab LQCD cluster, supported by
the Offices of Science, High Energy Physics, and Nuclear Physics of
the U.S. Department of Energy.  This research also used resources of
the National Energy Research Scientific Computing Center (NERSC),
which is supported by the Office of Science of the U.S. Department of
Energy under Contract No. DE-AC02-05CH11231 and resources of the
Argonne Leadership Computing Facility at Argonne National Laboratory,
which is supported by the Office of Science of the U.S. Department of
Energy under contract DE-AC02-06CH11357.

\bibliographystyle{apsrev4-1}
\bibliography{paper}{}

\appendix

\section{Ensembles studied}

\label{sec:data}

We use gauge field ensembles generated by the MILC collaboration
\cite{Bernard:2001av,Aubin:2004wf,Bazavov:2009bb} using $2+1$ flavors
of improved (asqtad) staggered sea quarks with various light quark
masses.  Relevant parameters of the gauge field ensembles in this
study are listed in Table~\ref{tab:ensembles}.  They fall into four
groups according to the approximate lattice spacing, namely coarse
(0.12 fm), fine (0.09 fm), superfine (0.06 fm), and ultrafine (0.045
fm).  The table shows the inverse lattice spacing in units of Sommer
parameter $r_0$.  The pion and $\bar s s$ pseudoscalar meson masses are
shown in lattice units.

\begin{table}[t]
\begin{tabular}{llllllr}
\hline
$10/g^2$& volume & $m_{ud}/m_s$ & $a m_\pi$ & $am_{\bar ss}$ & $r_0/a$ & $N_{\rm cfg}$\\
\hline
\multicolumn{7}{c}{coarse} \\
 6.85   & $20^3\times64$ & 0.05/0.05     & 0.48454(19) & 0.48454(19) &  3.921 & 364  \\
 6.83   & $20^3\times64$ & 0.04/0.05     & 0.43488(21) & 0.48647(22) &  3.889 & 340  \\
 6.79   & $20^3\times64$ & 0.02/0.05     & 0.31134(17) & 0.49012(18) &  3.860 & 469  \\
 6.76   & $20^3\times64$ & 0.01/0.05     & 0.22439(20) & 0.49427(18) &  3.822 & 644  \\
 6.76   & $20^3\times64$ & 0.007/0.05    & 0.18903(17) & 0.49324(16) &  3.847 & 435  \\
 6.76   & $24^3\times64$ & 0.005/0.05    & 0.15970(12) & 0.49261(14) &  3.865 & 317  \\
 8.00   & $20^3\times64$ & quenched      & $-$         & $-$         &  3.881 & 400  \\
\hline
\multicolumn{7}{c}{fine} \\
 7.18   & $28^3\times96$ & 0.031/0.031   & 0.32003(18) & 0.32003(18) &  5.580 & 447  \\
 7.11   & $28^3\times96$ & 0.0124/0.031  & 0.20638(18) & 0.32585(17) &  5.420 & 509  \\
 7.09   & $28^3\times96$ & 0.0062/0.031  & 0.14777(12) & 0.32698(8)  &  5.401 & 531  \\
 7.085  & $32^3\times96$ & 0.00465/0.031 & 0.12851(12) & 0.3269(2)   &  5.399 & 1000  \\
 7.08   & $40^3\times96$ & 0.0031/0.031  & 0.10538(6)  & 0.32744(8)  &  5.394 & 489  \\
 7.075  & $64^3\times96$ & 0.00155/0.031 & 0.0750(2)   & 0.3275(1)   &  5.398 & 890  \\
 8.40   & $28^3\times96$ & quenched      & $-$         & $-$         &  5.446 & 416  \\
\hline
\multicolumn{7}{c}{superfine} \\
 7.48   & $48^3\times144$ & 0.0072/0.018  & 0.13187(8)  & 0.20830(12) & 7.722 & 601  \\
 7.475  & $48^3\times144$ & 0.0054/0.018  & 0.11420(9)  & 0.2075(1)   & 7.722 & 618  \\
 7.47   & $48^3\times144$ & 0.0036/0.018  & 0.09353(6)  & 0.20731(6)  & 7.732 & 611  \\
 7.465  & $56^3\times144$ & 0.0025/0.018  & 0.07843(8)  & 0.20764(8)  & 7.726 & 518  \\
 7.46   & $64^3\times144$ & 0.0018/0.018  & 0.06678(3)  & 0.20749(4)  & 7.710 & 799  \\
 8.80   & $48^3\times144$ & quenched      & $-$         & $-$         & 7.388 & 405 \\
\hline
\multicolumn{7}{c}{ultrafine} \\
 7.81	& $64^3\times192$ & 0.0028/0.014  & 0.0712(1)   & 0.1583(1)   & 10.388 & 810  \\
\hline
\end{tabular}
\caption{Simulation parameters for the lattice ensembles used in this
  study, including measured masses of the Goldstone pion and $\bar s
  s$ meson, inverse lattice spacing in $r_0$ units, and number of
  configurations from the ensemble. For taste singlet pions, see Table
  \protect\ref{tab:split}.} \label{tab:ensembles}
\end{table}

\end{document}